\documentclass[11pt]{article}

\usepackage[left=1.5in,right=1.5in,top=1in,bottom=1in]{geometry}
\usepackage{graphicx}   
\usepackage{amsmath,amssymb}
\usepackage{cite}       
\usepackage{hyperref}   
\usepackage{caption}
\usepackage{float}

\title{Multimode fiber laser cavities as nonlinear optical processors}
\author{
Dilem E\c{s}lik$^{1}$, Bahad{\i}r Utku Kesgin$^{1}$, Fatma Nur K{\i}l{\i}n\c{c}$^{2}$ \and U\u{g}ur Te\u{g}in$^{1,2,*}$ \\
\small $^{1}$Department of Electrical and Electronics Engineering, Ko\c{c} University, \.{I}stanbul, T\"urkiye \\
\small $^{2}$Department of Computational Sciences and Engineering, Ko\c{c} University, \.{I}stanbul, T\"urkiye \\
\small $^{*}$Corresponding author: utegin@ku.edu.tr
}
\date{} 

\begin{document}
\maketitle

\begin{abstract}
Optical computing provides a promising path toward energy-efficient machine learning, yet implementing nonlinear transformations without complex electronics or high-power sources remains challenging. Here, we demonstrate that continuous-wave multimode fiber laser cavities can function as nonlinear optical processors. Input images encoded as phase patterns on a spatial light modulator undergo high-dimensional transformation through the interplay of multimode interference and gain saturation dynamics. The cavity maps input data into spatially stable, class-separable intensity distributions, enabling a simple linear classifier to achieve accuracies of 85--99\% across diverse benchmarks---including medical imaging and remote sensing---with orders of magnitude fewer trainable parameters than deep neural networks. Our results establish multimode fiber lasers as compact, low-power physical processors for scalable optical machine learning.
\end{abstract}

\section{Introduction}
The rapid growth of artificial intelligence has driven an unprecedented demand for computational resources~\cite{Wetzstein2020DeepOptics}. Training and deploying large-scale neural networks require extensive matrix operations and data movement, both of which consume significant power on conventional CMOS platforms due to charge transport and resistive dissipation~\cite{Miller2017Attojoule}. As model complexity continues to scale, the energy footprint of electronic accelerators poses a fundamental barrier, motivating the search for alternative computing paradigms~\cite{Miscuglio2020TensorCores}.

Optical computing offers a promising direction, as light propagation enables massively parallel information processing across spatial and spectral channels without resistive heating~\cite{Tegin2021SOLO,Pierangeli2019Ising,Wetzstein2020DeepOptics,Psaltis1990Holography}. Linear optical systems---including free-space diffractive networks~\cite{Lin2018Diffractive}, integrated photonic circuits~\cite{Shen2017Nanophotonic}, and fiber-based architectures~\cite{Kesgin2025FiberD2NN,Kesgin2025ChaosMMF}---can perform large-scale matrix operations at minimal energy cost~\cite{Hamerly2019LargeScale}. However, linear transformations alone are insufficient for general machine learning tasks, which inherently require nonlinear mappings between inputs and outputs~\cite{farhat1985optical}. Nonlinear operations enable expressive feature representations and class separation in high-dimensional spaces, forming the basis of modern neural network architectures~\cite{Appeltant2011Reservoir,Larger2017PhotonicReservoir}.

Introducing nonlinearity into optical systems typically requires electronic post-processing or specialized nonlinear media operating at high intensities, both of which compromise the energy and scalability advantages of photonic platforms~\cite{Tegin2021SOLO}. Reservoir computing offers an alternative framework, where a fixed high-dimensional dynamical system transforms inputs into linearly separable representations~\cite{rafayelyan2020large}. Speckle patterns generated by multimode interference have proven effective as random projection kernels for such tasks~\cite{Paudel2020Speckle}, and incorporating feedback through ring resonators further enriches the computational dynamics~\cite{Ashner2021Photonic}. Recent work has also demonstrated that nonlinear behavior can emerge from spatiotemporal instabilities in guided-wave systems~\cite{Kesgin2025ChaosMMF,wright2022nonlinear}, while programmable scattering media and broadband fiber dynamics provide additional mechanisms for nonlinear optical processing~\cite{Carpinlioglu2025GeneticONN,Maula2025Supercontinuum}.

Lasers inherently exhibit nonlinear dynamics through the interplay of gain, feedback, and saturation~\cite{Tegin2019SelfSimilar,turitsyn2011modeling}, and have been exploited for analog computation in coherent Ising machines and related architectures~\cite{mcmahon2016fully}. Single-mode laser reservoirs, however, are limited to a single complex degree of freedom~\cite{Brunner2013Reservoir}. Multimode fiber lasers, by contrast, support a large number of spatial modes whose competition for gain introduces rich, intensity-dependent dynamics~\cite{wright2015controllable}. When different input conditions are imposed on such a cavity, the resulting steady-state field distributions become input-dependent signatures shaped by modal interference and nonlinear gain saturation~\cite{Carpenter2015Modes,popoff2010measuring}.

Multimode fibers naturally provide high-dimensional optical representations: each spatial mode acts as an independent channel for encoding and transforming information~\cite{Rotter2017ComplexMedia,Ploschner2015MMF,Cao2023}. In a laser cavity, the number of supported modes determines the dimensionality of the feature space, while gain saturation reshapes this space into class-separable output patterns. The spatial modes thus serve as parallel computational channels, enabling physically embedded feature extraction without digital overhead.

Here, we demonstrate that continuous-wave multimode fiber laser cavities function as nonlinear optical processors for image classification. By encoding input images as phase patterns on a spatial light modulator, the cavity performs high-dimensional nonlinear feature transformation through combined multimode interference and gain saturation dynamics. The resulting steady-state intensity distributions serve as optical feature representations that a simple linear classifier can decode with high accuracy~\cite{Bueno2018Reinforcement,Antonik2019BrainInspired,McMahon2023PhysicsOC}. Across diverse benchmarks spanning natural images, aerial scenes, and medical diagnostics, our system achieves 85--99\% classification accuracy using orders of magnitude fewer trainable parameters than conventional deep networks. These results establish multimode fiber laser cavities as compact, low-power physical processors for scalable optical machine learning.

\section{Results}
\subsection{Numerical validation}

\begin{figure}[htbp]
\centering
\includegraphics[width=1\textwidth]{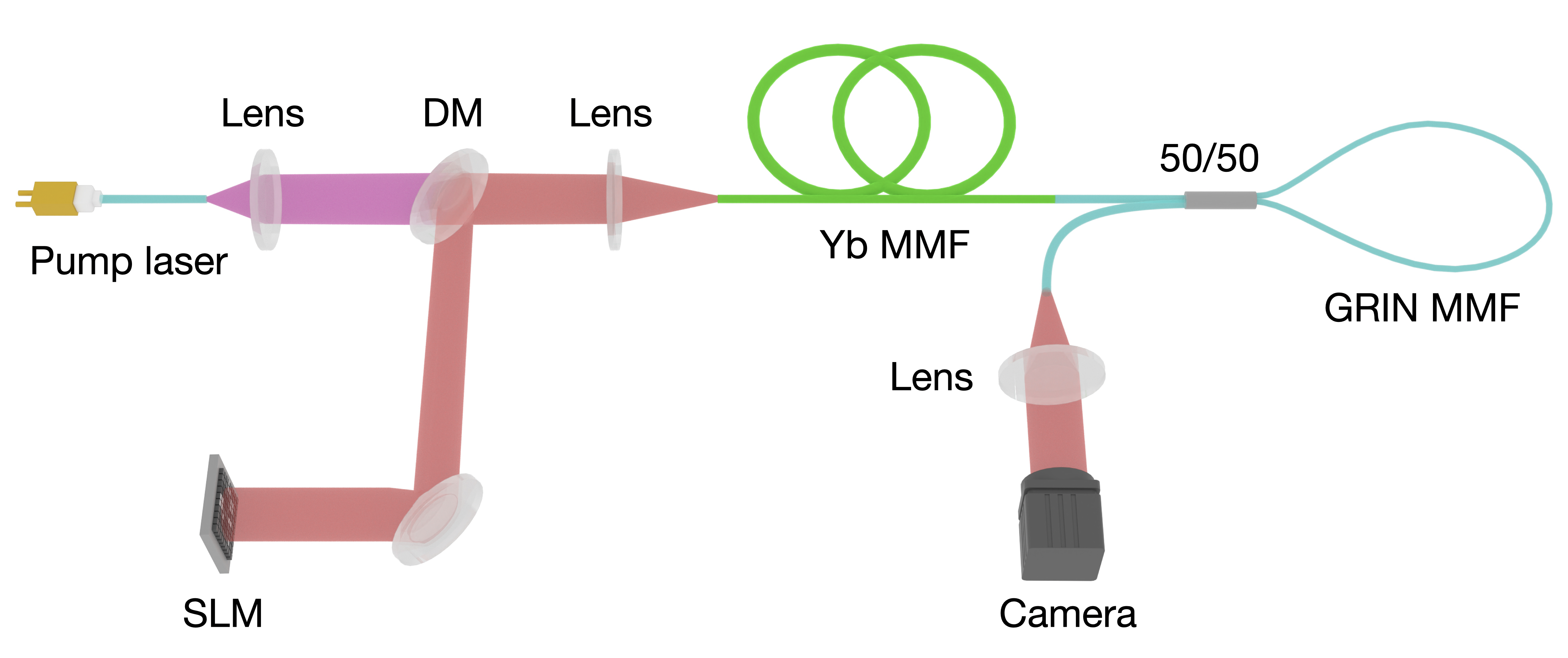}
\caption{Schematic of the multimode fiber laser computing platform. The cavity employs a linear geometry with a Sagnac loop, consisting of a Yb-doped gain fiber and a passive graded-index (GRIN) multimode fiber (MMF). A spatial light modulator (SLM) functions as a programmable end mirror to encode information, while a 50/50 coupler forms a feedback loop with output extraction.}
\label{fig1}
\end{figure}

Figure~\ref{fig1} illustrates our multimode fiber laser computing platform. The cavity comprises a spatial light modulator (SLM) serving as a programmable end mirror, a Yb-doped gain fiber, and a graded-index (GRIN) multimode fiber, with a 50/50 coupler forming a Sagnac loop for feedback and output extraction. To validate the concept, we first performed numerical simulations using the UCI Glass Identification benchmark~\cite{UCI_Glass}, which contains seven glass types characterized by nine compositional features.

\begin{figure}[htbp]
\centering
\includegraphics[width=\textwidth]{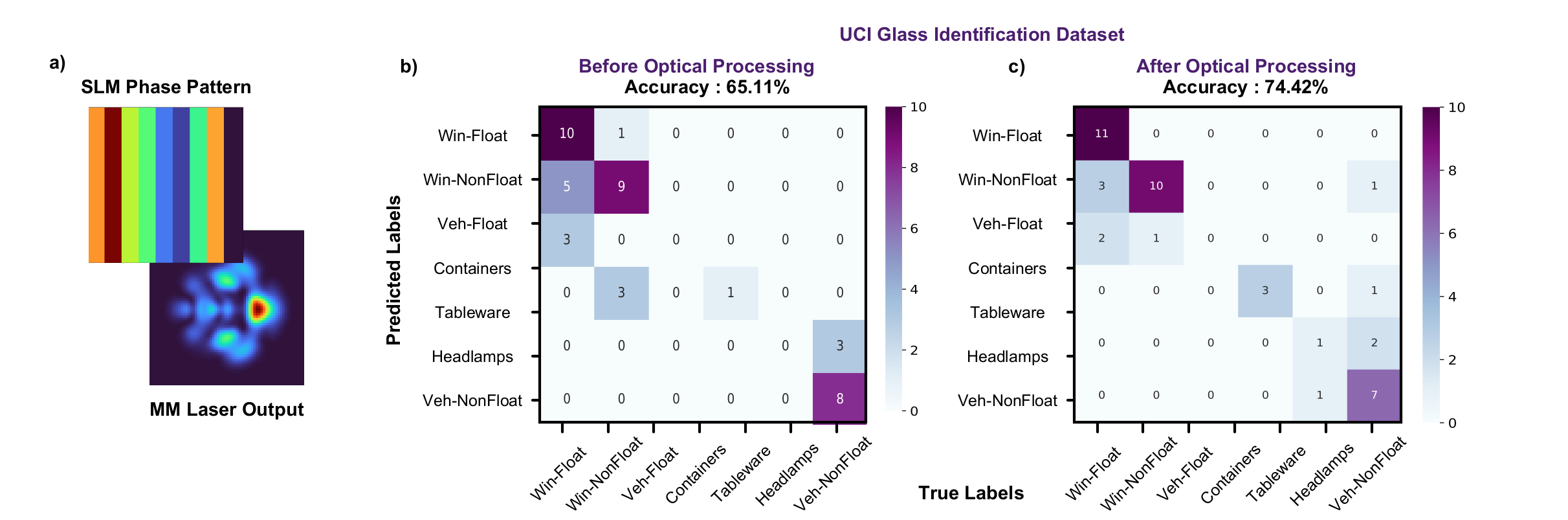}
\caption{Numerical simulation results. \textbf{a)} Representative SLM phase pattern (left) and corresponding laser output intensity (right). \textbf{b,c)} Confusion matrices for the UCI Glass dataset using raw features \textbf{(b)} and optically processed features \textbf{(c)}.}
\label{fig2}
\end{figure}

Each sample is encoded as a phase pattern on a $64\times64$ grid, where the nine features are mapped to vertical stripes of fixed width. The encoded beam propagates through the simulated cavity, undergoing diffraction, multimode interference, and saturable gain amplification. Representative input phase patterns and corresponding steady-state output intensity distributions are shown in Fig.~\ref{fig2}a. A linear Ridge classifier serves as the digital readout layer. Simulation details are provided in the Methods section and Supplementary Note~S2.

The numerical results confirm that cavity dynamics enhance classification performance. Using raw features, the linear classifier achieves 65.12\% accuracy (Fig.~\ref{fig2}b). After optical processing through the simulated laser cavity, accuracy improves to 74.41\% (Fig.~\ref{fig2}c). This enhancement arises from nonlinear gain saturation and multimode interference, which introduce class-dependent spatial structure that improves linear separability.

\subsection{Experimental demonstration}

Motivated by the numerical results, we constructed the fiber laser cavity shown in Fig.~\ref{fig1} and evaluated its performance on four image classification benchmarks: TrashNet~\cite{TrashNet}, RSSCN7~\cite{RSSCN7}, OCT MNIST~\cite{OCTMNIST}, and HAM10000~\cite{HAM10000}. These datasets span natural images, aerial scenes, and medical diagnostics, providing a diverse testbed for the optical processor. Input images are encoded as phase-only patterns on the SLM and processed by the multimode laser cavity. Dataset characteristics and preprocessing details are provided in Supplementary Note~S8.

When different phase masks are applied, the cavity converges to distinct, reproducible steady-state intensity distributions that depend sensitively on the input pattern. This behavior reflects input-dependent redistribution of power among spatial modes, driven by multimode interference and gain saturation. The cavity thus implements a deterministic nonlinear mapping from input phase patterns to stable spatial intensity signatures, which serve as optical feature representations for classification.

\begin{figure}[htbp]
\centering
\includegraphics[width=\textwidth]{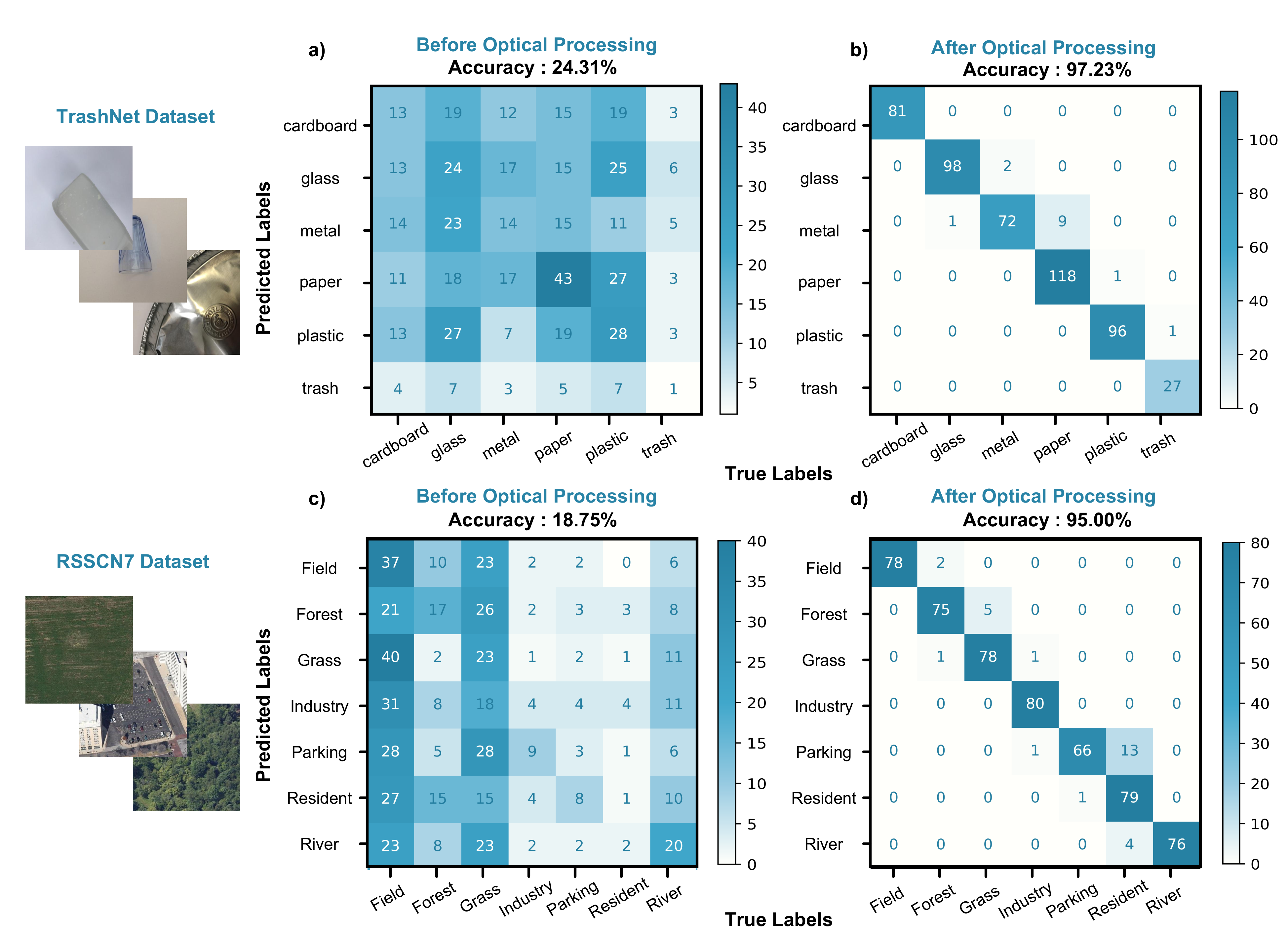}
\caption{Experimental classification results. \textbf{a,b)} Confusion matrices for TrashNet using raw images \textbf{(a)} and optically processed images \textbf{(b)}. \textbf{c,d)} Confusion matrices for RSSCN7 using raw images \textbf{(c)} and optically processed images \textbf{(d)}.}
\label{fig4}
\end{figure}

Across all datasets, optical processing yields substantial improvements over raw-image baselines evaluated under identical conditions (Supplementary Note~S7). For TrashNet, the system achieves 97.23\% accuracy with optically processed inputs, compared to 24.31\% for raw images (Fig.~\ref{fig4}a,b). For RSSCN7 aerial scene recognition, optical processing yields 95.00\% accuracy versus 18.75\% baseline (Fig.~\ref{fig4}c,d).

\begin{figure}[htbp]
\centering
\includegraphics[width=\textwidth]{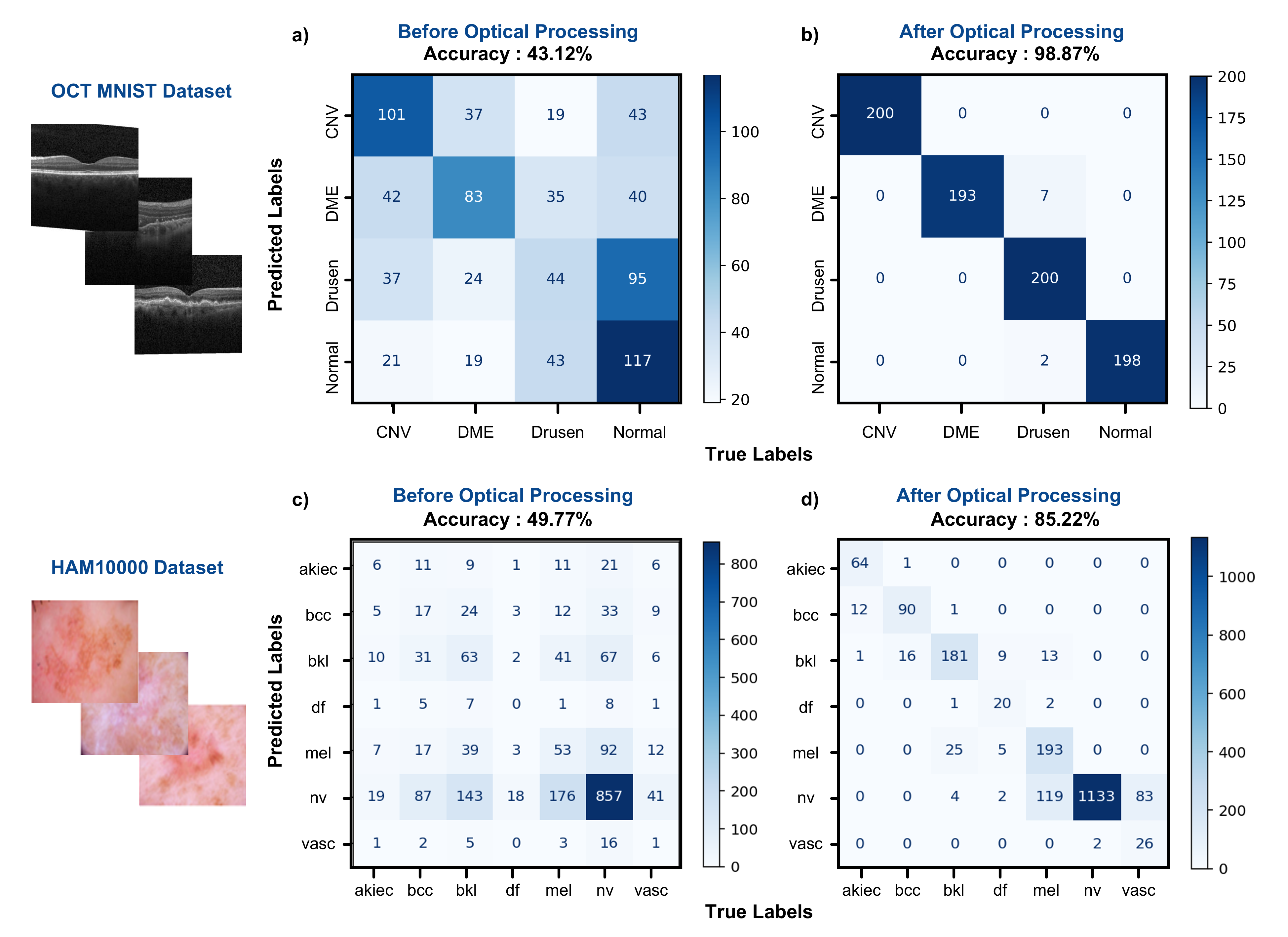}
\caption{Medical imaging classification results. \textbf{a,b)} Confusion matrices for OCT MNIST using raw images \textbf{(a)} and optically processed images \textbf{(b)}. \textbf{c,d)} Confusion matrices for HAM10000 using raw images \textbf{(c)} and optically processed images \textbf{(d)}.}
\label{fig3}
\end{figure}

Comparable improvements are observed for medical imaging tasks. On HAM10000, the optically processed outputs achieve 85.22\% accuracy compared to 49.77\% for raw images. For OCT MNIST (using a 4,000-image subset), the system reaches 98.78\% accuracy versus 43.12\% baseline. Confusion matrices for both medical datasets are shown in Fig.~\ref{fig3}.

\begin{figure}[H]
\centering
\includegraphics[width=\textwidth]{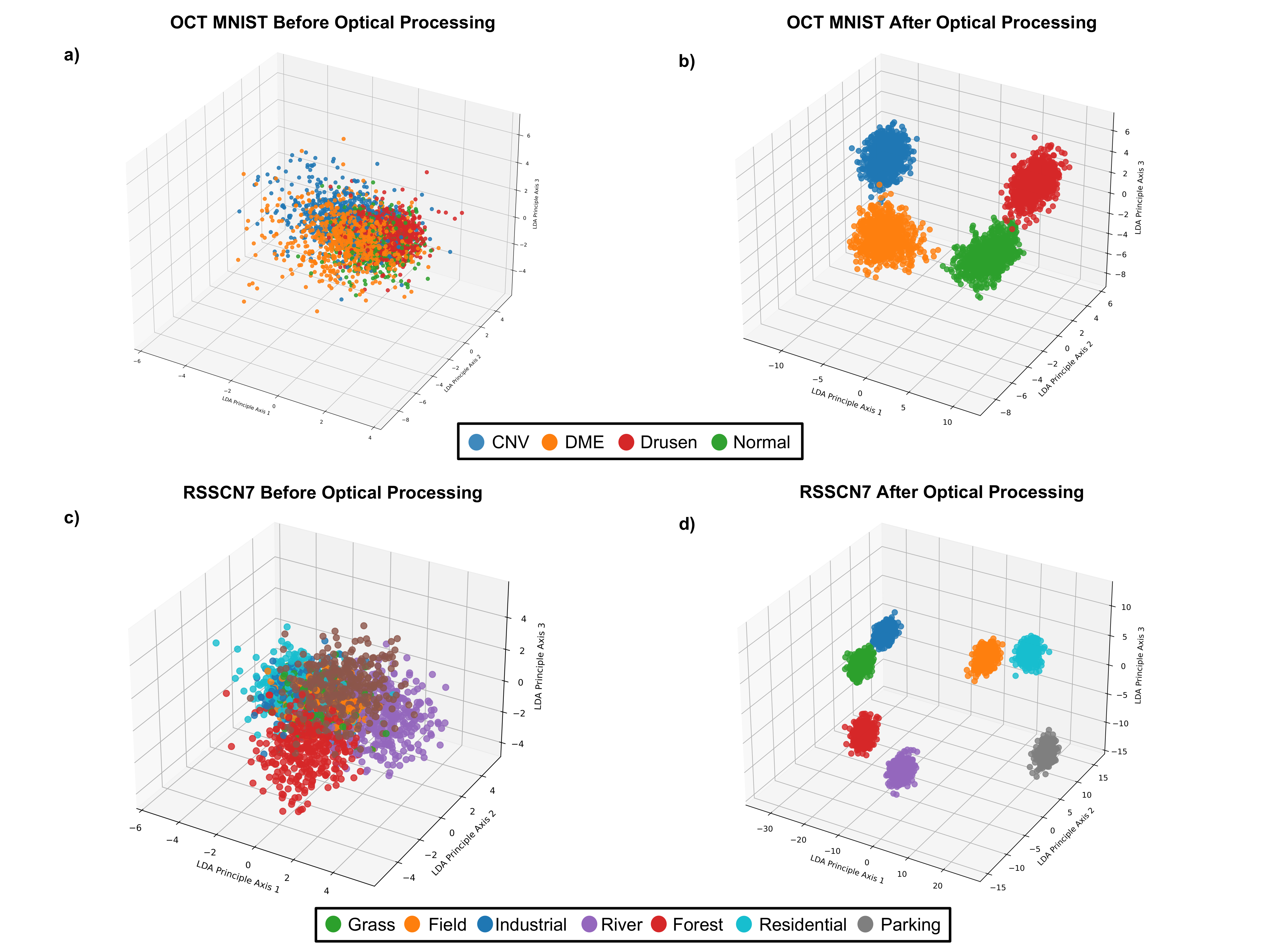}
\caption{Feature space visualization via Linear Discriminant Analysis. \textbf{a,b)} LDA projections for OCT MNIST using raw images \textbf{(a)} and optically processed images \textbf{(b)}. \textbf{c,d)} LDA projections for RSSCN7 using raw images \textbf{(c)} and optically processed images \textbf{(d)}.}
\label{fig5}
\end{figure}

To visualize the feature transformation, we applied Linear Discriminant Analysis (LDA) to both raw and optically processed data (Supplementary Note~S1). As shown in Fig.~\ref{fig5}, raw inputs exhibit significant class overlap, whereas optically processed features form well-separated clusters. Additional LDA visualizations are provided in Supplementary Figs.~S1--S2. This separation confirms that the multimode fiber laser acts as a physically implemented nonlinear feature extractor, mapping inputs into a high-dimensional space where classes become linearly separable.

Table~\ref{table2} compares the multimode laser (MML) computing approach with established convolutional neural networks. Despite using only a linear readout with 2,500--10,000 trainable parameters---several orders of magnitude fewer than deep networks---the MML system achieves competitive accuracy on both RSSCN7 and HAM10000. For fair comparison, the RSSCN7 results in Table~\ref{table2} are evaluated using a 50\%--50\% train--test split to match the CNN benchmarks; for HAM10000, all methods use consistent evaluation protocols.

\begin{table}[H]
\centering
\begin{tabular}{cccc}
\hline Learner & Parameter Count & RSSCN7 Accuracy* & HAM10000 Accuracy\\
\hline
\textit{\textbf{MML Computing}} & \textit{\textbf{2500 - 10000 }} & \textit{\textbf{94.21\%}} & \textit{\textbf{85.22\%}}\\
ResNet-50 & $26,000,000$  & $93.64\%$ & $85.30\%$ \\
VGG16 & $138,000,000$  &$93.57\%$ & $96.05\%$\\
AlexNet & $60,000,000$  &$91.85\%$ & $95.29\%$\\
\hline
\end{tabular}
\caption{Comparison of different machine learning models on the RSSCN7 and HAM10000 datasets. Benchmark accuracies for CNN-based models are taken from prior work~\cite{rsscn7benchref,ham10kbenchref}. For RSSCN7, all methods including MML use a 50\%--50\% train--test split for fair comparison. For HAM10000, consistent evaluation protocols are applied across all methods.}

\label{table2}
\end{table}

Taken together, the numerical simulations and experimental measurements consistently demonstrate that multimode fiber laser cavity dynamics significantly enhance the linear separability of diverse datasets. Across all tested tasks, the cavity maps input phase patterns into stable, high-dimensional spatial intensity distributions that enable accurate classification using a fixed, low-complexity linear readout. The agreement between simulation and experiment confirms that the observed performance gains arise from deterministic multimode interference combined with nonlinear gain saturation, rather than dataset-specific tuning or learned optical parameters.

\section{Discussion}
Our results demonstrate that multimode fiber laser cavities can serve as physically embedded nonlinear processors for machine learning tasks. The key mechanism is the interplay between multimode interference and gain saturation: input-dependent phase patterns excite different combinations of spatial modes, and the competition for gain reshapes these distributions into stable, class-discriminative steady states. This deterministic nonlinear transformation enables a simple linear classifier to achieve high accuracy without requiring trained optical elements or high-power operation.

The performance comparison in Table~\ref{table2} highlights a central advantage of this approach. While deep convolutional networks require tens to hundreds of millions of trainable parameters, our system achieves competitive accuracy using only 2,500--10,000 parameters in the linear readout layer. The physical cavity provides the nonlinear feature extraction that would otherwise require multiple layers of learned weights. This offloading of computation to the optical domain suggests a path toward hybrid optical--electronic systems where photonic preprocessing reduces the complexity and energy demands of digital backends.

Several features distinguish the multimode laser cavity from other optical computing approaches. Compared with passive multimode fiber systems~\cite{Paudel2020Speckle}, intracavity gain amplifies the nonlinear response and ensures reproducible steady-state outputs. Unlike free-space diffractive networks~\cite{Lin2018Diffractive}, the fiber geometry provides a compact, alignment-stable platform. The continuous-wave operating regime avoids the complexity and power requirements of pulsed nonlinear systems, while the Sagnac loop configuration ensures stable feedback without sensitive interferometric alignment.

Several limitations should be noted. The current implementation processes one image at a time, with throughput limited by SLM refresh rates (typically 30--60~Hz). The system also requires calibration to account for environmental perturbations to the fiber. While classification accuracy is competitive with deep networks on the tested benchmarks, performance on more challenging tasks---such as fine-grained recognition or adversarial robustness---remains to be evaluated. The comparison with CNNs in Table~\ref{table2} should also be interpreted cautiously, as train--test protocols differ across studies.

Looking forward, several directions could extend the capabilities of this platform. Integration with faster spatial light modulators or digital micromirror devices could increase throughput by orders of magnitude. Scaling to fibers with larger mode counts would expand the dimensionality of the optical feature space. Task-specific optimization of cavity parameters---such as fiber length, gain level, and coupling ratios---could further enhance performance for targeted applications. On-chip implementations using multimode waveguides and integrated gain sections offer a path toward compact, manufacturable photonic processors. These developments could position multimode laser cavities as practical components in energy-efficient optical computing architectures.

\section{Methods}

\subsection{Cavity model and simulation}
The intracavity optical field is modeled using a (2+1)D beam propagation framework based on the split-step Fourier method~\cite{Agrawal2020NFO}. The slowly varying field envelope $A(x,y,z)$ evolves under combined diffraction and nonlinear gain according to

\begin{equation}
    \frac{\partial A}{\partial z} = \frac{i}{2k_0}(\frac{\partial^2}{\partial x^2} +\frac{\partial^2 }{\partial y^2})A - \frac{ik_0 \Delta n}{R^2} A + \frac{g}{2}A + i\gamma |A|^2A
\end{equation}

where refractive index $\Delta n$ and gain $g$ terms which vary according to fiber type can be expressed as:

\begin{equation}
    \Delta n (x,y)=
    \begin{cases}
    \Delta, \quad  \text{if step index and } (x^2+y^2) < R_{core} \\
    \Delta (x^2+y^2), \quad \text{if graded index} \\
    0,\quad \text{if} \quad (x^2+y^2) > R_{core}
    \end{cases}
\end{equation}

\begin{equation}
    g(x,y)=
    \begin{cases}
    \frac{\log_{10}(gss)}{l_{fiber}}\times \frac{1}{1+\frac{|A(x,y)|^2}{I_{sat}}}, \quad  \text{if doped fiber } \\
    0, \quad \text{if non-doped fiber}
    \end{cases}
\end{equation}

Here $\Delta$ is the relative index difference and $gss$ refers to maximum linear gain out of the fiber. Under continuous-wave operation, temporal dispersion is negligible, and the dominant nonlinear mechanism is gain saturation in the Yb-doped fiber: modes with higher local intensity experience stronger gain depletion, leading to input-dependent redistribution of modal power. Instantaneous Kerr nonlinearities are neglected at the moderate intracavity intensities used here. Steady-state solutions are obtained by iterating the cavity propagation over multiple round trips until the intensity distribution stabilizes. Numerical implementation details and, boundary conditions are provided in Supplementary Notes~S2--S3.

\subsection{Experimental setup}
The experimental system is a continuous-wave Yb-doped multimode fiber laser operating in a linear cavity geometry with a Sagnac loop (Fig.~\ref{fig1}). The gain section consists of a 1~m step-index Yb-doped fiber (20/125~$\mu$m core/cladding), pumped at 976~nm with 1.2~W optical power. The passive section uses a 4~m graded-index multimode fiber (50/125~$\mu$m), with 2~m incorporated in the Sagnac loop. A 50/50 fiber coupler forms the loop mirror, providing stable feedback and 50\% output extraction.

Phase modulation is implemented using a spatial light modulator (Holoeye Pluto 2.1; 1920$\times$1080 pixels; 8~$\mu$m pixel pitch) acting as a programmable cavity end mirror. Steady-state output intensity patterns are captured using a monochrome camera (FLIR BFS-U3-04S2M-CS). Further experimental details are provided in Supplementary Note~S6.

\subsection{Data encoding and classification}
Input images are converted to grayscale and resized to a uniform resolution before encoding as phase-only patterns on the SLM. The applied phase pattern sets the transverse phase distribution of the intracavity field, determining the initial modal excitation at the fiber input. The resulting steady-state output intensity distributions serve as optical feature representations.

Classification is performed using a linear Ridge classifier with $\ell_2$ regularization. Data are split into training and test sets using stratified sampling (80\%--20\% for most experiments). Feature extraction, pooling procedures, and classifier implementation details are provided in Supplementary Notes~S4--S8.

\subsection*{Data Availability}
Data underlying the figures are publicly available at Zenodo \cite{eslik_2026_18495686}. Any additional data generated during experiments may be obtained from the authors upon reasonable request. Datasets used in the study are publicly available through their respective access links.

\subsection*{Code Availability}
Code of the numerical validation is publicly available at Zenodo\cite{eslik_2026_18495686}. Any other code related to the results in this work may be obtained from the authors upon reasonable request.

\subsection*{Acknowledgements}
This work is supported by the Scientific and Technological Research Council of Turkey (TUBITAK) under grant number 122C150.

\subsection*{Author Contributions}

D.E. designed and performed the experiments, built the optical setup, and carried out the experimental measurements.
B.U.K. developed and performed the numerical simulations.
F.N.K. contributed to the data encoding with spatial light modulator (SLM).
U.T. conceived and supervised the project.
All authors discussed the results and contributed to the manuscript.

\subsection*{Competing Interests}
U.T. is an Editorial Board Member for \textit{Communications Physics}, but was not involved in the editorial review of, or the decision to publish this article. All other authors declare no competing interests.

\bibliographystyle{naturemag}
\bibliography{manuscript}

\end{document}